\documentclass[12pt]{article}
\usepackage[dvips]{graphics,color}
\usepackage{graphicx}
\usepackage{dcolumn}
\usepackage{bm}
\usepackage{epsfig}

\newcommand{\beq}{\begin{equation}}
\newcommand{\eeq}{\end{equation}}
\newcommand{\beqa}{\begin{eqnarray}}
\newcommand{\eeqa}{\end{eqnarray}}
\newcommand{\bd}[1]{ \mbox{\boldmath $#1$}}

\begin{document}
\def\ii{\'\i}

\title{
The Robertson-Walker Metric in
a Pseudo-complex General Relativity
}
\author{Peter O. Hess$^{1}$, Leila Maghlaoui$^{2}$ and Walter Greiner$^{2}$ \\
{\small\it
$^1$Instituto de Ciencias Nucleares, UNAM, Circuito Exterior, C.U.,}\\ 
{\small\it A.P. 70-543, 04510 M\'exico D.F., Mexico} \\
{\small\it $^2$Frankfurt Institute for Advanced Studies, Johann Wolfgang Goethe Universit\"at,} \\
{\small\it Ruth-Moufang-Str. 1, 60438 Frankfurt am Main, Germany} \\
}

\maketitle

\abstract{
We investigate the consequences of the
pseudo-complex General Relativity within a pseudo-complexified
Roberston-Walker metric.
A contribution to the energy-momentum tensor arises, which corresponds
to a dark energy and may change with the radius of the universe, i.e., time.
Only when the Hubble function $H$ does not change in time, the solution is
consistent with a constant $\Lambda$.
}


\vskip 1cm
\section{Introduction}

In the past several paths have been proposed on how to extend the theory of
{\it General Relativity} (GR).
One of the first was Einstein himself \cite{einstein1,einstein2} who,
in an attempt to unify electrodynamics with GR, extended GR to
{\it complex} GR. For more
recent articles on complex GR, you may consult \cite{mantz,lovelook}.
Others \cite{caneloni,brandt1,brandt2,brandt3,beil1,beil2,beil3,beil4,moffat1,moffat2,kunstatter}
introduced a maximal acceleration, related to a minimal length parameter
in the theory, 
or \cite{crumeyrolle1,crumeyrolle2,clerc1,clerc2} tried to extend GR by introducing
hyperbolic coordinates.

The introduction of a maximal acceleration
and the use of hypercomplex (a synonym for para-complex)
coordinates is very much related to what was
published in \cite{hess}.
In \cite{hess} a pseudo-complex ({\it pc}) extension of
General Relativity (GR) was proposed.
(To the article \cite{hess} we will refer from here on as (I).)
The main objective
was to investigate the analogue of the Schwarzschild metric
and its consequences within the pseudo-complex extension of GR. Its
possible experimental verification was and is of prime importance.
The {\it pc}-GR is formulated in a different manner than in
\cite{crumeyrolle1,crumeyrolle2,clerc1,clerc2}. Because pseudo-complex
numbers exhibit a zero divisor basis (see section II), we were able to define
two independent theories of GR in each zero-divisor component, which were later connected. The resulting length squared element is similar but not
equal to 
\cite{caneloni,brandt1,brandt2,brandt3,beil1,beil2,beil3,beil4,moffat1,moffat2,kunstatter}.
It contains an additional term proportional to the velocity times
acceleration.
As already said, the analogue of the Schwarzschild solution was studied within the new formalism.
A strong heuristic assumption
was made, requiring that the {\it pc} scalar curvature ${\bd {\cal R}}=0$.
This did lead to strong corrections, which are excluded by the
{\it Parametrized-Post-Newton formalism} (PPN) \cite{misner}.
However, relaxing the
condition gives corrections which are smaller and not contradicting
the PPN formalism; details will be given in a forthcoming publication.
Deviations of the redshift to the standard Schwarzschild case were calculated.

One of the main results and messages in (I) was: The pseudo-complex
description, with its extended variational principle, contains no singularity, i.e. black holes don't  exist. Furthermore, the new theory (I) automatically
introduces dark energy, whose density depends in the analogue of the Schwarzschild case
on the radial distance. In fact, the increase of the dark energy density towards
smaller radial distances finally hinders (prevents) the collapse of a large mass.
The increase of dark energy, as a function of the radial
distance, depends sensitively on some functions $\xi_k$, which are very
difficult to deduce. The results are in line with the model described in \cite{mich1} where the
dark energy is described by a scalar field coupled to gravity. It is interesting and quite helpful for understanding, to compare our results with those obtained within this model. In the model \cite{mich1} the
relativistic equations are solved numerically. As a result, dark energy
accumulates around the central mass, reducing the radius
of the event horizon. The advantage
of this model \cite{mich1} is that it provides a distribution of the scalar field, whose
intensity should be proportional to our $\xi$ functions. In one case, the
dark energy falls off very quickly with the radial distance. The disadvantage
is the large numerical effort needed. The similarity between this procedure
and our theory, however, indicates that
both investigations might be useful for each other.

Here, in this contribution, we intend to show that the same effect introduces a
dark energy in models of the Robertson-Walker (RW) type universes
\cite{adler,peebles}. Our theory also contains a minimal length scale.
Its influence on solutions might be important for large mass concentrations
\cite{feoli}. Here, we will not discuss it but refer to a later
publication.
Whether our theory is realized in nature also depends on
the predictions which can be made and on their experimental
verification. In the present contribution our principal objective is
to identify some consequences of the pseudo-complex theory,
and to determine whether possible differences to Einsteins General Relativity may be observed.

In section II a short review on how to define {\it pc} variables
and their properties will be given. This is for those readers who
are not yet accustomed to this mathematical structure.
A simple presentation can also be found in (I).
In section III we will formulate the
{\it pc} extension of the Robertson-Walker metric. It will be quite
straight forward and the steps can be copied from \cite{adler} or any
other book on GR may be consulted.
For details, we will give mainly reference to
\cite{adler}.
In section IV we will solve the {\it pc}-RW model and in section V we shall
discuss some consequences. Section VI summarizes the main conclusions.

\section{Pseudo-complex variables and pseudo-complex metric}

Here we give a brief resum\'e on pseudo-complex variables,
helpful to understand various steps presented in this contribution.
The formulas, presented here, can be used without going into the details.
A more profound introduction to pseudo-complex variables is given in
\cite{field1,field2}, which may be consulted for better understanding.

The pseudo-complex variables are also known as
{\it hyperbolic} \cite{crumeyrolle1,crumeyrolle2},
{\it hypercomplex} \cite{kantor} or {\it para-complex} \cite{galea}.
We will continue to use the term pseudo-complex.

The pseudo-complex variables are {\it defined} via

\beqa
X & = & x_1 + I x_2 ~~~,
\eeqa
with $I^2=1$. This is similar to the common complex notation except for
the different behavior of $I$. An alternative presentation is to introduce
the operators

\beqa
\sigma_\pm & = & \frac{1}{2} \left( 1 \pm I \right)  \nonumber \\
\eeqa
with
\beqa
\sigma_\pm^2 & = & \sigma_\pm ~~~,~~~ \sigma_+\sigma_- = 0 ~~~.
\eeqa
The $\sigma_\pm$ form a so called {\it zero divisor basis},
with the zero divisor defined in mathematical terms by
$\bd{P}^0 = \bd{P}^0_+ \cup \bd{P}^0_-$, with
$\bd{P}^0_\pm=\left\{ X=\lambda \sigma_\pm| \lambda ~\epsilon~ \bd{R} \right\}$.
The zero divisor generates all the differences to the complex number.

This basis is used to rewrite the pseudo-complex variables as

\beqa
X & = & X_+ \sigma_+ + X_- \sigma_- ~~~,
\eeqa
with

\beqa
X_\pm & = & x_1 \pm x_2  \nonumber \\
{\rm or} \nonumber \\
x_1 & = & \frac{1}{2}\left( X_+ + X_- \right)
~~~,~~~ x_2  ~=~  \frac{1}{2} \left( X_+ - X_- \right)
~~~.
\eeqa

The pseudo-complex conjugate of a pseudo-complex variable is

\beqa
X^* & = & x_1 - I x_2 = X_+\sigma_- + X_- \sigma_+ ~~~.
\eeqa
The {\it norm} square of a pseudo-complex variable is given by

\beqa
|X|^2 = XX^ * & = & x_1^2 - x_2^2 ~~~=~~~ X_+X_- ~~~.
\eeqa
This allows for the appearance of a positive, negative and {\it null norm}.
Variables with a zero norm are members of the zero-divisor, i.e.,
they are either proportional to $\sigma_+$ or $\sigma_-$.

{\it It is very useful to carry out all calculations within the zero divisor
basis, $\sigma_\pm$. Here, all manipulations can be realized
independently in both sectors, because $\sigma_+\sigma_-=0$.}

In each zero divisor component,
differentiation and multiplication can be manipulated in
the same way as with real or complex variables.
For example, we have \cite{field2}

\beqa
F(X) & = & F(X_+) \sigma_+ + F(X_-) \sigma_-
\label{f1}
\eeqa
and a product of two functions $F(X)$ and $G(X)$ satisfies

\beqa
& F(X)G(X) &
\nonumber \\
& = \left( F(X_+)\sigma_+ + F(X_-)\sigma_- \right)
\left( G(X_+)\sigma_+ + G(X_-)\sigma_- \right) &
\nonumber \\
& = F(X_+)G(X_+) \sigma_+ + F(X_-)G(X_-) \sigma_-  &
~~~,
\label{fg1}
\eeqa
because $\sigma_+ \sigma_- = 0$ and $\sigma_\pm^2 = \sigma_\pm$.
As a further example, we have

\beqa
\frac{F(X)}{G(X)} & = & \frac{F(X_+)}{G(X_+)}\sigma_+ +
\frac{F(X_-)}{G(X_-)}\sigma_-
~~~.
\label{eq10}
\eeqa
This can be proved as follows:

\beqa
& \frac{F(X)}{G(X)}= \frac{ F(X_+)\sigma_+ + F(X_-)\sigma_-}
{G(X_+)\sigma_+ + G(X_-)\sigma_- } &
\nonumber \\
& = \frac{ \left( F(X_+)\sigma_+ + F(X_-)\sigma_-\right)
\left( G(X_+)\sigma_- + G(X_-)\sigma_+\right)}
{\left( G(X_+)\sigma_+ + G(X_-)\sigma_-\right)
\left( G(X_+)\sigma_- + G(X_-)\sigma_+\right)} &
\eeqa
where we have multiplied the numerator and denominator by the
pseudo-complex conjugate of $G(X)$, using $\sigma_+^*=\sigma_-$.
With $\sigma_+\sigma_-=0$ and $\sigma_\pm^2=\sigma_\pm$,
the last expression can be written as

\beqa
& \frac{ \left( F(X_+)G(X_-) \sigma_+ + F(X_-)G(X_+)\sigma_-\right)}
{G(X_+)G(X_-)\left( \sigma_+ + \sigma_- \right)}
~~~.
\eeqa
Because $\sigma_+ + \sigma_- = 1$, we arrive at Eq. (\ref{eq10}).

Differentiation is defined as

\beqa
\frac{DF(X)}{DX} & = & \lim_{\Delta X \rightarrow 0}
\frac{F(X+\Delta X)-F(X)}{\Delta X}
~~~,
\eeqa
where $D$ refers from here on to the pseudo-complex infinitesimal
differential.

A very important difference to the standard GR is the introduction
of a {\it modified} variational principle. It states that the variation of an
action has to be within the zero-divisor, i.e.

\beqa
\delta S & \epsilon & {\bd {\cal P}}^0
~~~,
\label{variation}
\eeqa
where ${\bd {\cal P}}^0$ denotes the zero divisor given by all values which
are either proportional to $\sigma_+$ ($\lambda \sigma_+$) or
to $\sigma_-$ ($\lambda \sigma_-$). For convenience, the latter is chosen
in this contribution, as it was in (I). However, instead of
"$\epsilon~ {\cal P}^0$" we write here "$=\lambda \sigma_-$".
The number at the right hand side of
(\ref{variation}) has zero norm and can be treated as a
"generalized zero", thus representing a minimal extension of the
variational principle.

In the {\it pc}-GR the metric is pseudo-complex, i.e., (see (I) for details)

\beqa
g_{\mu\nu} & = & g^+_{\mu\nu} \sigma_+ + g^-_{\mu\nu} \sigma_- ~~~.
\label{metric}
\eeqa
Each component ($\sigma_\pm$) can be treated independently, for many
purposes, e.g., as how to define parallel displacement, Christoffel symbols,
etc. Most of the steps,
known from standard GR can be
carried out analogously. This is the advantage of using the
zero-divisor basis. It can also be shown that the four-divergence of
the metric tensor is zero (see (I)).

The connection of both components of the zero-divisor basis
happens through the variation
principle, mentioned above. It states that, with the convention used and
$\delta S=\delta S_+ \sigma_+ + \delta S_- \sigma_-$, we have

\beqa
\delta S_+ & = & 0 ~~~{\rm and}~~~ \delta S_- ~=~ \lambda
~~~.
\eeqa
For the length element square

\beqa
d\omega^2 & = & g_{\mu\nu} DX^\mu DX^\nu \nonumber \\
& = & g^+_{\mu\nu} DX_+^\mu DX_+^\nu \sigma_+
+ g^-_{\mu\nu} DX_-^\mu DX_-^\nu \sigma_+
\nonumber \\
& = & DX_\mu^+DX^\mu_+ \sigma_+ + DX_\mu^-DX^\mu_- \sigma_-
~~~,
\eeqa
reality is imposed through

\beqa
d\omega^{*2} & = & d\omega^2
~~~.
\eeqa
This implies that $d\omega^{2}$ is equal to its pseudo real part

\beqa
d\omega^{2} & = &
\frac{1}{2}
\left( DX_\mu^+ DX^\mu_+ + DX_\mu^- DX^\mu_- \right)
\nonumber \\
& = & dx_\mu dx^\mu + l^2 du_\mu du^\mu 
\label{omegaR}
\eeqa
where in the last step we simply substituted $X_\mu^\pm$ and
$X^\mu_\pm$ by their expressions in terms of the coordinates and four-velocities.
The pseudo-imaginary component has to vanish, i.e.,

\beqa
0 & = & \frac{1}{2}\left( DX_\mu^+ DX^\mu_+ - DX_\mu^- DX^\mu_- \right)
\nonumber \\
& = & l\left( dx_\mu du^\mu + du_\mu dx^\mu \right)
~~~.
\label{omegaI}
\eeqa
One way to proceed is to take (\ref{omegaR}) as the final length element
and impose (\ref{omegaI}) as a constriction, noting that (\ref{omegaI})
gives the dispersion relation: Integrating, this gives
$u_\mu u^\mu = 1$. This is a nice feature because in other models
the dispersion relation is usually put in by hand.

Using Eq. (37) of (I), which
reads in differential form

\beqa
dx_\mu & = & g^0_{\mu\nu} dx^\nu + lh_{\mu\nu} du^\nu   \nonumber \\
ldu_\mu & = & lg^0_{\mu\nu} du^\nu + h_{\mu\nu} dx^\nu
~~~,
\label{relations}
\eeqa
the length element ({\ref{omegaR}) and the constriction (\ref{omegaI})
can be respectively written as

\beqa
d\omega^2 & = & g^0_{\mu\nu} ( dx^\mu dx^\nu + l^2du^\mu du^\nu )
\nonumber \\
&& + lh_{\mu\nu} (dx^\mu du^\nu + du^\mu dx^\nu )
~~~,
\nonumber \\
\label{domega2}
\eeqa
and

\beqa
& h_{\mu\nu} \left(dx^\mu dx^\nu + l^2 du^\mu du^\nu \right) &
\nonumber \\
& +lg^0_{\mu\nu} \left(dx^\mu du^\nu + du^\mu dx^\nu \right)  =  0 &
~~~.
\eeqa
These are the expressions reported in (I).

Some models \cite{mantz,lovelook}, using complex coordinates,
start with a real length element squared,
defining it as $\frac{1}{2}\left[ d\omega^* + d\omega \right]$.
If one does the same with pseudo-complex coordinates, one misses
then the simplifications involved, treating first the zero divisor
components separately. In the complex version \cite{mantz,lovelook} there
is no zero divisor and, therefore, one does not have this advantage.

\section{Pseudo-complex Robertson-Walker Metric}

Previously we followed the steps in 
chapter 12.3 of the book of Adler-Bazin-Schiffer \cite{adler}. Of course any
other book on GR can be consulted. In this section we
shall mainly repeat these steps, for
sake of completeness, with the difference that the variables are now
pseudo-complex. The reader will see that the formulation is identical to
standard GR, with the difference of the appearance of additional functions
due to the modified variational principle.

In order to proceed, one choses so called {\it Gaussian coordinates}, in
which one uses a distinguished (absolute) time coordinate, thus the abandonment of a
completely covariant treatment of the cosmological problem \cite{adler}.
This is the price one has to pay to simplify the cosmological models
and to describe physical reality in convenient mathematical terms. Such
coordinates were first introduced by Gauss within a different context.

The pseudo-complex length element in Gaussian coordinates,
before imposing reality, is given by

\beqa
d\omega^2 & = & (dX^0)^2 - e^{G(X^0,R)} \left( dR^2 + R^2 d\theta^2
+ R^2 sin^2\theta d\phi^2 \right) \nonumber \\
& = & (dX^0)^2 - e^{G(X^0,R)}d\Sigma^2
~~~,
\label{line}
\eeqa
where we already used the pseudo-complex coordinates. $G$ is a function
of time and the radial coordinate $R$.
As will now be shown, $G$
can be written as the sum of the functions $g(X^0)$ and $f(R)$, the
first one depending only on time and the second one only on $R$, i.e.,
{\bf $G(X^0,R)=g(X^0) + f(R)$}.
The starting point is the equivalence principle that two observers at
two different points observe the same physics. The only difference may be
in the scale the two observers use. Thus the ratio of the proper distance
element at two different space points $R_1$ and $R_2$ must remain
fixed in time:

\beqa
\frac{e^{G(X^0,R_1)}}{e^{G(X^0,R_2)}} & = & {\rm const~in~time} ~~~,
\label{eq23}
\eeqa
i.e., this ratio must be independent of $X^0$. Therefore one must have

\beqa
G(X^0,R_1) & = & G(X^0,R_2)+ F(R_1,R_2) ~~~,
\eeqa
which then yields for Eq. (\ref{eq23})

\beqa
\frac{e^{G(X^0,R_2)+ F(R_1,R_2)}}{e^{G(X^0,R_2)}} & = &
e^{F(R_1,R_2)}
~~~,
\eeqa
which is independent on time.

If we choose a fixed value for $R_2$ we can write

\beqa
G(X^0,R_1) & = & g(X^0)+ f(R_1) ~~~.
\eeqa

\begin{center}
{\bf Christoffel symbols}:
\end{center}

The equation for the geodesics is given by

\beqa
\delta\int\left[ ({\dot X}^0)^2 -e^G \left( {\dot R}^2
+R^2 {\dot \theta}^2 +R^2 sin^2\theta {\dot \phi}^2\right)\right]ds
& \epsilon & {\bd P}^0
~~~,
\nonumber \\
\eeqa
with $s$ being a curve parameter. In addition,
we used the new variational procedure,
requiring that the variation gives a number within the zero-divisor basis.

After variation, the following equations of motion are obtained
(a dot refers to the derivation with respect to $s$, the curve parameter,
and a prime indicates for
the function $g$ a derivative with respect to $X^0$ while for $f$ it is
a derivative with respect to $R$).

\beqa
& {\ddot X}^0 + \frac{1}{2}g^\prime e^G \left( {\dot R}^2 + R^2{\dot \theta}^2
+ R^2 sin^2\theta {\dot \phi}^2 \right) & =  \xi_0 \sigma_-
\nonumber \\
&{\ddot R} + \frac{1}{2} f^\prime {\dot R}^2 + g^\prime {\dot X}^0 {\dot R}
&
\nonumber \\
&-\left( \frac{1}{2} f^\prime + \frac{1}{R} \right)
\left( R^2{\dot \theta}^2
+ R^2 sin^2\theta {\dot \phi}^2 \right) & =  \xi_R \sigma_-
\nonumber \\
&{\ddot \theta} + 2\left( \frac{1}{2} f^\prime + \frac{1}{R} \right)
{\dot R}{\dot \theta} + g^\prime {\dot X}^0{\dot \theta}
- sin\theta cos\theta{\dot \phi}^2 & = \xi_\theta \sigma_-
\nonumber \\
&{\ddot \phi} + 2\left( \frac{1}{2} f^\prime + \frac{1}{R} \right)
{\dot R}{\dot \phi}
+ g^\prime {\dot X}^0{\dot \phi} + 2 {\dot \theta}{\dot \phi}cot\theta
& =  \xi_\phi \sigma_-
~~~.
\eeqa
We used the convention that on the right hand side stands an element
in the zero divisor basis proportional to $\sigma_-$. Choosing it proportional
to $\sigma_+$ would give equivalent results, i.e., the $\sigma_-$ and
$\sigma_+$ components are just interchanged.

Comparing this with the equation of motion

\beqa
{\ddot X}^\mu +
\left\{
\begin{array}{ccc}
& \mu & \\
\nu && \lambda
\end{array}
\right\}
{\dot X}^\nu {\dot X}^\lambda & \epsilon & {\cal P}^0 ~~~,
\label{eq-m-x}
\eeqa
yields the non-zero Christoffel symbols (others can be deduced using
the symmetry properties of the Christoffel symbols):

\beqa
\left\{
\begin{array}{ccc}
& 0 & \\
1 && 1
\end{array}
\right\} &=& \frac{1}{2}g^\prime e^G
\nonumber \\
\left\{
\begin{array}{ccc}
& 0 & \\
2 && 2
\end{array}
\right\} &=& \frac{1}{2}g^\prime e^G R^2
\nonumber \\
\left\{
\begin{array}{ccc}
& 0 & \\
3 && 3
\end{array}
\right\} &=& \frac{1}{2}g^\prime e^G R^2 {\rm sin}^2\theta
\nonumber \\
\left\{
\begin{array}{ccc}
& 1 & \\
0 && 1
\end{array}
\right\} &=& \frac{1}{2}g^\prime
\nonumber \\
\left\{
\begin{array}{ccc}
& 1 & \\
1 && 1
\end{array}
\right\} &=& \frac{1}{2}f^\prime
\nonumber \\
\left\{
\begin{array}{ccc}
& 1 & \\
2 && 2
\end{array}
\right\} &=& -R^2\left( \frac{1}{2}f^\prime + \frac{1}{R} \right)
\nonumber \\
\left\{
\begin{array}{ccc}
& 1 & \\
3 && 3
\end{array}
\right\} &=& -R^2\left( \frac{1}{2}f^\prime + \frac{1}{R} \right)
{\rm sin}^2\theta
\nonumber \\
\left\{
\begin{array}{ccc}
& 2 & \\
0 && 2
\end{array}
\right\} &=& \frac{1}{2}g^\prime 
~=~
\left\{
\begin{array}{ccc}
& 3 & \\
0 && 3
\end{array}
\right\}
\nonumber \\
\left\{
\begin{array}{ccc}
& 2 & \\
1 && 2
\end{array}
\right\} &=&
\left( \frac{1}{2}f^\prime + \frac{1}{R} \right)
~=~
\left\{
\begin{array}{ccc}
& 3 & \\
1 && 3
\end{array}
\right\}
\nonumber \\
\left\{
\begin{array}{ccc}
& 2 & \\
3 && 3
\end{array}
\right\} &=& -{\rm sin}\theta {\rm cos}\theta
\nonumber \\
\left\{
\begin{array}{ccc}
& 3 & \\
2 && 3
\end{array}
\right\} &=& {\rm cot}\theta
~~~.
\label{chris}
\eeqa
>From the line element we find for the determinant of the metric tensor

\beqa
{\rm ln}\sqrt{-g} & = & \frac{3}{2}g(X^0) + \frac{3}{2}f(R) + 2 {\rm ln}R
+ {\rm ln}\mid {\rm sin}\theta\mid ~~~.
\eeqa
Using the Christoffel symbols given in (\ref{chris}), one finds

\beqa
\left\{
\begin{array}{ccc}
& \mu & \\
0 && 0
\end{array}
\right\}_{\mid \mu} &=& 0
\nonumber \\
\left\{
\begin{array}{ccc}
& \mu & \\
1 && 1
\end{array}
\right\}_{\mid \mu} &=& \frac{1}{2} e^G \left( g^{\prime\prime}
+ g^{\prime 2} \right) + \frac{1}{2} f^{\prime\prime}
\nonumber \\
\left\{
\begin{array}{ccc}
& \mu & \\
2 && 2
\end{array}
\right\}_{\mid \mu} &=&
\left[ \frac{1}{2} e^G \left( g^{\prime\prime} + g^{\prime 2} \right)
\right. \nonumber \\
&& \left. 
-\left( \frac{1}{2}f^{\prime\prime} + \frac{1}{R} f^\prime
+\frac{1}{R^2} \right)\right] R^2
\nonumber \\
\left\{
\begin{array}{ccc}
& \mu & \\
3 && 3
\end{array}
\right\}_{\mid \mu} &=&
\left[ \frac{1}{2} e^G \left( g^{\prime\prime} + g^{\prime 2} \right)
\right. \nonumber \\
&& \left.
-\left( \frac{1}{2}f^{\prime\prime} + \frac{1}{R} f^\prime
\right)\right] R^2{\rm sin}^2\theta 
\nonumber \\
&& - {\rm cos}^2\theta
~~~.
\eeqa
The following relations are also useful

\beqa
\left\{
\begin{array}{ccc}
& \mu & \\
0 && \nu
\end{array}
\right\}
\left\{
\begin{array}{ccc}
& \nu & \\
0 && \mu
\end{array}
\right\}
 &=&   \frac{3}{4} g^{\prime 2}
\nonumber \\
\left\{
\begin{array}{ccc}
& \mu & \\
1 && \nu
\end{array}
\right\}
\left\{
\begin{array}{ccc}
& \nu & \\
1 && \mu
\end{array}
\right\}
& = & \frac{1}{2}e^G g^{\prime 2} + \frac{3}{4} f^{\prime 2}
+ \frac{2}{R}f^\prime + \frac{2}{R^2}
\nonumber \\
\left\{
\begin{array}{ccc}
& \mu & \\
2 && \nu
\end{array}
\right\}
\left\{
\begin{array}{ccc}
& \nu & \\
2 && \mu
\end{array}
\right\}
& = & \left[ \frac{1}{2}e^G g^{\prime 2} - \frac{1}{4} f^{\prime 2}
- \frac{2}{R}f^\prime 
\right. \nonumber \\
&& \left.
- \frac{2}{R^2} + \frac{1}{R^2} {\rm cot}^2\theta \right]
R^2
\nonumber \\
\left\{
\begin{array}{ccc}
& \mu & \\
3 && \nu
\end{array}
\right\}
\left\{
\begin{array}{ccc}
& \nu & \\
3 && \mu
\end{array}
\right\}
& = & \left[ \frac{1}{2}e^G g^{\prime 2} - \frac{1}{2} f^{\prime 2}
- \frac{2}{R}f^\prime 
\right. \nonumber \\
&& \left.
- \frac{2}{R^2} + \frac{1}{R^2} {\rm cot}^2\theta \right]
\nonumber \\
&& \times R^2 {\rm sin}^2\theta
~~~.
\eeqa

With this, the non-vanishing components of the Ricci tensor are
(after some calculations using also the definition of the Ricci tensor in
terms of the Christoffel symbols, identical to standard GR)

\beqa
{\cal R}_{00} & = & \frac{3}{2}g^{\prime\prime} + \frac{3}{4}g^{\prime 2}
\nonumber \\
{\cal R}_{11} & = & f^{\prime\prime} + \frac{1}{R}f^\prime
- e^G \left( \frac{1}{2}g^{\prime\prime} + \frac{3}{4}g^{\prime 2} \right)
\nonumber \\
{\cal R}_{22} & = & \left[
\frac{1}{2}f^{\prime\prime} + \frac{1}{4} f^{\prime 2}
+ \frac{3}{2R} f^\prime
- e^G \left( \frac{1}{2}g^{\prime\prime} + \frac{3}{4}g^{\prime 2} \right)
\right] R^2
\nonumber \\
{\cal R}_{33} & = & \left[
\frac{1}{2}f^{\prime\prime} + \frac{1}{4} f^{\prime 2}
+ \frac{3}{2R} f^\prime
- e^G \left( \frac{1}{2}g^{\prime\prime} + \frac{3}{4}g^{\prime 2} \right)
\right] 
\nonumber \\
&& \times R^2 {\rm sin}^2\theta
~~~.
\eeqa
All other components are zero.

{\it The last equations were just copied from} \cite{adler}.

To obtain the tensor component ${\cal R}^\mu_\nu$ we need the
expression for the metric tensor and its inverse. We have

\beqa
g_{\mu\nu} & = & \left(
\begin{array}{cccc}
1 & 0 & 0 & 0 \\
0 & -e^G & 0 & 0 \\
0 & 0 & -e^G R^2 & 0 \\
0 & 0 & 0 & -e^G R^2 sin^2 \theta
\end{array}
\right)                       
\label{metric3}
\eeqa
and
\beqa
g^{\mu\nu} & = & \left(
\begin{array}{cccc}
1 & 0 & 0 & 0 \\
0 & -e^{-G} & 0 & 0 \\
0 & 0 & -\frac{e^{-G}}{R^2} & 0 \\
0 & 0 & 0 & -\frac{e^{-G}}{R^2 sin^2\theta} 
~~~.
\end{array}
\right)                       
\label{metric4}
\eeqa

With this we get (${\cal R}_\mu^\nu = g^{\nu\rho}{\cal R}_{\mu\rho}$)

\beqa
{\cal R}^0_0 & = & \frac{3}{2}g^{\prime\prime} + \frac{3}{4}g^{\prime 2}
\nonumber \\
{\cal R}^1_1 & = & \left( \frac{1}{2}g^{\prime\prime}
+ \frac{3}{4}g^{\prime 2} \right)
- e^{-G} \left( f^{\prime\prime} + \frac{f^\prime}{R} \right)
\nonumber \\
{\cal R}^2_2 & = & {\cal R}^3_3 ~=~
\left( \frac{1}{2}g^{\prime\prime}
+ \frac{3}{4}g^{\prime 2} \right)
\nonumber \\
&&- e^{-G} \left( \frac{1}{2}f^{\prime\prime} + \frac{1}{4} f^{\prime 2}
+ \frac{3f^\prime}{2R} \right)
~~~.
\nonumber \\
\eeqa
The Riemann curvature is then

\beqa
{\cal R} & = & 3\left( g^{\prime\prime} + g^{\prime 2}\right)
-2e^{-G} \left( f^{\prime\prime} + \frac{f^{\prime 2}}{4}
+\frac{2}{R} f^\prime \right)
\eeqa

Denoting the energy momentum tensor by $T^\mu_\nu$ and exploiting the above results, the equations of motion are

\beqa
-\frac{8\pi\kappa}{c^2} T^0_0 & = &
\left[ e^{-G}\left( f^{\prime\prime} + \frac{f^{\prime 2}}{4}
+\frac{2f^\prime}{R} \right) - \frac{3}{4}g^{\prime 2} \right]
+ \xi_0\sigma_-
\nonumber \\
-\frac{8\pi\kappa}{c^2} T^1_1 & = &
\left[ e^{-G}\left( \frac{f^{\prime 2}}{4}
+\frac{f^\prime}{R} \right) - g^{\prime\prime} -
\frac{3}{4}g^{\prime 2} \right]
+ \xi_1\sigma_-
\nonumber \\
-\frac{8\pi\kappa}{c^2} T^2_2 & = &
\left[ e^{-G}\left( 
\frac{f^{\prime\prime}}{2} +  \frac{f^{\prime}}{2R}\right)
- g^{\prime\prime} - \frac{3}{4}g^{\prime 2} \right]
+ \xi_2\sigma_-
\nonumber \\
-\frac{8\pi\kappa}{c^2} T^3_3 & = &
\left[ e^{-G}\left( 
\frac{f^{\prime\prime}}{2} + \frac{f^{\prime}}{2R}\right)
- g^{\prime\prime} - \frac{3}{4}g^{\prime 2} \right]
+ \xi_3\sigma_-
\nonumber \\
-\frac{8\pi\kappa}{c^2} T^\mu_\nu & = & 0 ~~,~~ \mu\neq\nu
~~~.
\label{motion-0}
\eeqa
The $f^\prime (R)$ refers to the derivative with respect to $R$,
while $g^\prime (X^0)$ refers to the derivative with respect to
$X^0$.
The $\xi_\mu$ functions appear due to the new variational principle.
In \cite{adler} there appears instead the cosmological
constant $\Lambda$. In principle we can add such a constant, too.
However, one of the reasons not to do so, is that the $\xi$
functions will reproduce such an effect. Inspecting Eq. (\ref{motion-0})
one can identify $\xi_k$ as additional diagonal contributions to the
energy-momentum tensor, times $\sigma_-$. 
Transferring $\xi_0$ in the first equation of (\ref{motion-0}) to the left hand side and
factorizing $\frac{8\pi\kappa}{c^2}$, we can associate to $\xi_0$
the energy density

\beqa
\rho_\Lambda & = & \frac{c^2}{8\pi\kappa}\xi_0 \sigma_-
~~~.
\label{rho-lambda}
\eeqa
Later, when we project to the pseudo-real part, this will give a contribution
to the pseudo-real energy density, associated to dark energy.
This fact will be useful in
understanding the results obtained further below.

\section{Solving the Equations of Motion}

Homogeneity of the matter distribution requires that

\beqa
& T^1_1 ~=~ T^2_2 ~=~ T^3_3 & ~~~.
\label{homog}
\eeqa
Due to the similarity to the $T_k^k$,
the same argument can be used for the $\xi$ functions, giving

\beqa
& \xi_1 ~=~ \xi_2 ~=~ \xi_3 & ~~~.
\label{homog-2}
\eeqa
The $\xi_k$ ($k=1,2,3$) may be functions of the time.

Taking an appropriate difference of the equation for $T^2_2$ with
$T^1_1$, leads to the equation (subtract second equation in
(\ref{motion-0}) from the third one in (\ref{motion-0}))

\beqa
f^{\prime\prime} - \frac{1}{2}(f^\prime )^2 - \frac{f^\prime}{R} & = & 0
~~~,
\eeqa
i.e, {\it the same} equation as given in \cite{adler}. The solution is
also supplied (as proposed in \cite{adler} by direct insertion):

\beqa
e^f & = & \frac{b^2}{\left[ 1 - \frac{ab}{4}R^2 \right]^2} ~~~,
\eeqa
with $a$ and $b$ as constants. Finally, the length square
element takes the form,
after some redefinitions ($\mid ab \mid = \frac{1}{R_0^2}$ \cite{adler})

\beqa
d\omega^2 & = & (dX^0)^2 - e^{g(X^0)} \frac{1}
{\left( 1+\frac{kR^2}{4R_0^2} \right)^2} d\Sigma^2 ~~~.
\eeqa
It is exactly of the same form as in standard GR, with the difference
that the coordinates are now pseudo-complex.
This is in distinction to the pseudo-complex Schwarzschild metric
\cite{hess}, where the differences appear already in the functional
form of the metric. The $k$ acquires the values
$k=0$, $\pm 1$, while $R_0^2$ is a constant, related to $a$ and $b$ via
$\mid ab \mid = 1/R^2_0$. The function $g(X^0)$ is yet undetermined.
The $k$-values of 0, $\pm$1 can be used to model different universes.
However, as we know now, the $k=0$ case is
the only one which is consistent with experiment \cite{peebles}. This is the case
we will finally study explicitly. It may serve as an example for
other studies.

We will now use the co-moving pseudo-complex
coordinates \cite{adler}, i.e., ${\dot X}^0=1$ and ${\dot X}^1$
= ${\dot X}^2$ = ${\dot X}^3$ = 0, where the dot refers to the
derivative with respect to the eigen-time.

The energy-momentum tensor takes the form

\beqa
\left( T^\mu_\nu \right) & = &
\left(
\begin{array}{cccc}
\rho &&& \\
& -\frac{p}{c^2} && \\
&& -\frac{p}{c^2} & \\
&&& -\frac{p}{c^2}  \\
\end{array}
\right) ~~~,
\label{energy-momentum}
\eeqa
which is quite standard.
The $\rho$ is the matter density and $p$ the pressure,
assumed here to be pseudo-real, though, in general they can be
pseudo-complex.
{\it We nevertheless will already
restrict to pseudo-real functions, as they should be}.

The relevant functions in the length element take the form (see Eq. (13.3)
of \cite{adler})

\beqa
e^{G(X^0,R)} & = & \frac{{\bd R}(X^0)^2}{R_0^2 \left( 1+kR^2/(4R_0^2) \right)^2}
\nonumber \\
e^{g(X^0)} & = & {\bd R}(t)^2
\nonumber \\
e^{f(R)} & = & \frac{1}{R_0^2 \left( 1+kR^2/(4R_0^2) \right)^2}
~~~,
\label{eq46}
\eeqa
which are directly obtained from the expression of the
length square element,
with some redefinitions. The ${\bd R}$ should not be confused with the
pseudo-complex radius variable $R$, nor with the Riemann curvature
${\cal� R}$. ${\bd R}$ is an object with a length unit
and is interpreted as the pseudo-complex {\it radius of the universe}.

{\it From now on, let us substitute $X^0$ by its pseudo-real part
$ct$. For example ${\bd R}(X^0)$ will be written as ${\bd R}(t)$.
The derivative
with respect to $X^0$ is converted into a derivation with respect to
$ct$, i.e., $\frac{d{\bd R}}{d(ct)}$ = $ \frac{1}{c}\frac{d{\bd R}}{dt}$
= $\frac{{\bd R}^\prime}{c}$.}

We part now from the above equations of motion (\ref{motion-0}).
The expressions in the functions $f$ and $g$ and their derivatives can
be re expressed in terms of the variable ${\bd R}$ using (\ref{eq46}).
For example $f=-{\rm ln}R_0^2 - 2{\rm ln}\left(1+kR^2/(4R_0^2)\right)$ and
$g=2{\rm ln}{\bd R}$. This and their derivatives have to be inserted into
(\ref{motion-0}).
Using the
symmetry conditions of homogeneity (\ref{homog})
and (\ref{homog-2})  and the form of the
energy-momentum tensor (\ref{energy-momentum}),
the equations of motion acquire the form (remember that
${\bd R}^\prime=\frac{d{\bd R}}{dt}$)

\beqa
\frac{8\pi\kappa}{c^2} \rho & = & -\xi_0\sigma_-
+\left[ \frac{3k}{{\bd R}(t)^2} + \frac{3}{c^2} \frac{{\bd R}^\prime (t)^2}
{{\bd R}(t)^2} \right]
\nonumber \\
\frac{8\pi\kappa}{c^2} \frac{p}{c^2} & = & \xi_1\sigma_-
-\left[ \frac{k}{{\bd R}(t)^2} + \frac{{\bd R}^\prime (t)^2}
{c^2{\bd R}(t)^2} + \frac{2{\bd R}^{\prime\prime}(t)}{c^2{\bd R}(t)} \right]
~~~.
\label{eq11}
\eeqa
The details of intermediate steps can be directly copied from any
book on General Relativity, e.g. \cite{adler}.
Instead of four equations we have only two, due to the symmetry conditions
(\ref{homog})and (\ref{homog-2}).
The prime refers now to the derivative with respect to the time $t$.

{\it Here, we will obtain one of our results}, which can be extracted without
detailed knowledge of the $\xi_k$ functions. It will be
useful in order to understand the ansatz in the relation between
$\xi_1$ and $\xi_0$ further below (see Eqs. (\ref{bedi}) and (\ref{eq17})):
Assuming that the density
$\rho$ and the pressure $p$ are pseudo-real quantities, the
$\sigma_-$ component of Eq. (\ref{eq11}) tells us that

\beqa
\xi_0 & = &
\frac{3k}{{\bd R}_-(t)^2} + \frac{3}{c^2} \frac{{\bd R}_-^\prime (t)^2}
{{\bd R}_-(t)^2}
\nonumber \\
\xi_1 & = & 
\frac{k}{{\bd R}_-(t)^2} + \frac{{\bd R}_-^\prime (t)^2}
{c^2{\bd R}_-(t)^2} + \frac{2{\bd R}_-^{\prime\prime}(t)}{c^2{\bd R}_-(t)}
~~~.
\label{x0-x1}
\eeqa
Here ${\bd R}_-(t)$ is the $\sigma_-$ component of the radius of the universe.
In order to get an idea what this implies for $\xi_0$ and $\xi_1$,
without having to solve the problem, we can use the experimental result
(${\bd R}_r$ is the pseudo-real component of the universe)

\beqa
\frac{{\bd R}_r^\prime}{{\bd R}_r} & = & {\rm H} ~~~{\rm with}~~~
{\rm H}^\prime <<1
~~~,
\eeqa
where the prime refers to the derivative with respect to time and
${\rm H}$ is the Hubble constant. Because ${\bd R}={\bd R}_r + l {\bd R}_I$,
${\bd R}_I$ being the pseudo-imaginary component of ${\bd R}$,
and $l$ is the length parameter of the theory, which is extremely small
(see (I)), we also can assume that ${\bd R} \approx {\bd R}_r$ and, because
$R_I=\frac{1}{2}\left(R_+-R_- \right)$, we can set
${\bd R}_\pm = \approx {\bd R}_r$. Using this we can
approximately write, assuming a nearly constant
$H=\frac{{\bd R}_r^\prime}{{\bd R}_r}$,

\beqa
\frac{{\bd R}_r^{\prime\prime}}{{\bd R}_r} & = &
\frac{{\bd R}_r^{\prime\prime}}{{\bd R}_r^\prime}
\frac{{\bd R}_r^{\prime}}{{\bd R}_r} ~~~=~~~
({\rm ln}{\bd R_r}^\prime  )^\prime \frac{{\bd R}_r^{\prime}}{{\bd R}_r}
~~~=~~~ \left[ {\rm ln} ({\rm H}{\bd R}_r) \right]^\prime
\frac{{\bd R}_r^{\prime}}{{\bd R}_r}
\nonumber \\
& = &  \left[ {\rm ln}{\rm H} + {\rm ln}{\bd R}_r \right]^\prime
\frac{{\bd R}_r^{\prime}}{{\bd R}_r}
\nonumber \\
& \approx & \left( \frac{{\bd R}_r^{\prime}}{{\bd R}_r}  \right)^2
~=~ H^2
~~~.
\label{estimation}
\eeqa

With that, utilizing (\ref{x0-x1}) and $k=0$, we can write the $\xi_0$ and $\xi_1$ approximately as

\beqa
\xi_0 & \approx & \frac{3}{c^2} {\rm H}^2 ~~~ \approx ~~~ \xi_1
~~~.
\label{bedi}
\eeqa
Further below we will see that this exactly corresponds to the case of
a cosmological constant not changing with the redshift.
This would be an exact result, if ${\rm H}$
is constant all over the history of the universe.
Knowing that ${\rm H}$ is changing in time, implies
that there {\it must} be a dependence on the radius of the universe, i.e.,
the redshift $z$.
For that we have to solve
the equation of motion exactly. As we will see further below,
it is not easy to get the exact form of $\xi_0$ and $\xi_1$ but rather the
use of a parametrization is appropriate.

{\it Let us now continue to solve the {\it pc}-RW model:}

Taking again appropriate linear combinations of (\ref{eq11})
(the first equation of (\ref{eq11}) plus three times the second equation
of (\ref{eq11}) and the first equation of (\ref{eq11}) plus
the second one), we get new versions of the form

\beqa
\frac{4\pi\kappa}{c^2} \left( \rho + \frac{3p}{c^2} \right) & = &
\frac{1}{2} \left( 3\xi_1 - \xi_0 \right) \sigma_-
- \frac{3{\bd R}^{\prime\prime}}
{c^2 {\bd R}}
\nonumber \\
\frac{4\pi\kappa}{c^2} \left( \rho + \frac{p}{c^2} \right) & = &
\frac{1}{2} \left( \xi_1 - \xi_0 \right)\sigma_- 
+ \frac{k}{{\bd R}^2}
+ \frac{{\bd R}^{\prime 2}
-{\bd R}{\bd R}^{\prime\prime}}{c^2 {\bd R}^2}
~~~.
\nonumber \\
\label{eq-mot}
\eeqa
Using that

\beqa
\frac{{\bd R}{\bd R}^{\prime\prime}-{\bd R}^{\prime 2}}{c^2 {\bd R}^2} & = &
\frac{d}{dt} \left(\frac{{\bd R}^\prime}{c^2{\bd R}} \right) ~~~,
\eeqa
we arrive at the equation

\beqa
\frac{d}{dt} \left(\frac{{\bd R}^\prime}{c^2{\bd R}}\right)  & = &
\frac{1}{2}\left( \xi_1 - \xi_0 \right) \sigma_- +\frac{k}{{\bd R}^2}
- \frac{4\pi\kappa}{c^2} \left( \rho + \frac{p}{c^2} \right)
~~~.
\label{eqa}
\eeqa

Differentiation of the first equation in (\ref{eq11}) with respect to time
gives

\beqa
\frac{8\pi\kappa}{c^2} \frac{d\rho}{dt} & = &
-\frac{d\xi_0}{dt}\sigma_- - \frac{6k}{{\bd R}^3}\bd R^\prime
+ \frac{6{\bd R}^\prime}{\bd R} \frac{d}{dt}
\left( \frac{1}{c^2}\frac{{\bd R}^\prime}{\bd R} \right) ~~~.
\eeqa
Substituting (\ref{eqa}) into (58) and multiplying the result
by $\frac{c^2}{8\pi\kappa} {\bd R}^3$, yields

\beqa
{\bd R}^3 \frac{d\rho}{dt} & = & \frac{c^2}{8\pi\kappa}
\left[ -{\bd R}_-^3 \frac{d\xi_0}{dt} + 3{\bd R}_-^2{\bd R}_-^\prime
(\xi_1 - \xi_0) \right] \sigma_- \nonumber \\
& & -3{\bd R}^2{\bd R}^\prime (\rho + \frac{p}{c^2} )
~~~.
\eeqa
Note that $3{\bd R}^2{\bd R}^\prime$ = $\frac{d{\bd R}^3}{dt}$. Shifting the
last term of this equation to the left hand side leads to

\beqa
\frac{d}{dt} \left( \rho {\bd R}^3 \right)
+ \frac{p}{c^2} \frac{d{\bd R}^3}{dt} & = &
\frac{c^2}{8\pi\kappa}
\left[ \frac{d{\bd R}_-^3}{dt} \left( \xi_1 - \xi_0 \right)
- {\bd R}_-^3 \frac{d\xi_0}{dt} \right]\sigma_-
~~~.
\nonumber \\
\label{eq15}
\eeqa
Identifying the mass within a given volume of the universe by $M=\rho V$,
with $V$ as a given volume, the last equation can be written as

\beqa
\frac{dM}{dt} + \frac{p}{c^2} \frac{dV}{dt} & = &
\frac{c^2}{8\pi\kappa}
\left[ \frac{dV_-}{dt} \left( \xi_1 - \xi_0 \right)
- V_- \frac{d\xi_0}{dt} \right]\sigma_-
~~~.
\label{eq16}
\eeqa
{\it This is a local energy balance!} In order to maintain local energy
conservation, we have to require that the right hand side is zero.
This leaves us with the condition

\beqa
\frac{d\xi_0}{dt} & = & \frac{d(lnR_-^3)}{dt} \left( \xi_1 - \xi_0 \right)
~~~.
\label{eq17}
\eeqa
{\it Any solution for $\xi_0$ and $\xi_1$ has to fulfill this differential
equation.} The negative index of $V$ refers to the fact that the
equation holds in the $\sigma_-$ component. Using $\xi_1 = \xi_0$ leads to
$\frac{d\xi_0}{dt}=0$, or $\xi_0=\xi_1=\Lambda ={\rm const}$. I.e., for this
case we recover the model with a cosmological constant not changing with
time. This equation is not sufficient to solve for $\xi_0$
and $\xi_1$; in fact one condition is missing.

The first equation in (\ref{eq-mot}) has the usual interpretation when
$\xi_0$ = $\xi_1$ = 0. Then the left hand side is the sum of
two positive quantities, the density and the pressure. The right hand side
of (\ref{eq-mot})
is proportional to the acceleration ${\bd R}^{\prime\prime}$
of the radius of the universe,
${\bd R}$, multiplied by (-1).
This equation tells us that the acceleration of ${\bd R}$
has to be negative, i.e., we get a de-acceleration. {\it In contrast, in the
pseudo-complex description there is an additional term}
$\frac{1}{2} \left( 3\xi_1 - \xi_0 \right)\sigma_-$
{\it present, which might be positive}. Transferring
it to the left hand side may give in total a negative function in time,
i.e., {\it depending of the functional form of} $\xi_0$ {\it and}
$\xi_1$ {\it in time, an accelerated phase may be reproduced or not}.

{\bf Let us see whether we can get also acceleration, i.e., that
${\bd R}^{\prime\prime}>0$ in the pseudo-complex version of GR}:

Using the left hand side of Eq. (\ref{eq15}) (the right hand side is
set to zero as argued below Eq. (\ref{eq16})),
we obtain, after multiplying with $dt$,

\beqa
{\bd R}^3 d\rho + 3{\bd R}^2\rho d{\bd R} + \frac{p}{c^2}3{\bd R}^2d{\bd R}
& = & 0
~~~.
\eeqa
Dividing by $3{\bd R}^3$ we obtain

\beqa
\frac{d\rho}{3} + \left( \rho + \frac{p}{c^2} \right)
\frac{d{\bd R}}{{\bd R}} & = & 0
~~~.
\eeqa
Finally, dividing by $(\rho + \frac{p}{c^2})$ yields

\beqa
\frac{d\rho}{3\left( \rho + \frac{p}{c^2} \right)}
+\frac{d{\bd R}}{{\bd R}} & = & 0
~~~.
\label{rho-R}
\eeqa

Now we have to make an assumption on the {\it equation of state}! {\bf This
is a delicate part} and the results can change, depending on
which equation of state we take. The equation of state may
also depend on different time epochs.
The basic assumptions are that i) the distribution of the mass in the
universe can be treated as an ideal gas, dust or radiation,
the mass being equally
distributed (this is only approximately true).
The equation of state is

\beqa
p & = & \alpha \rho ~~~,
\eeqa
where $\rho$ is the {\it energy density} and $\alpha$ is zero for a model
with dust, $\frac{2}{3}$ for a classical ideal gas and $\frac{1}{3}$ for
a relativistic ideal gas (radiation).

With this, (\ref{rho-R}) can be solved with the solution

\beqa
\rho & = & \rho_0 R^{-3(1+\frac{\alpha}{c^2})} ~~~,
\label{rho}
\eeqa
where the $\rho_0$ is a pseudo-complex integration constant. Its dimension is
density.

This result is substituted into the first equation of (\ref{eq-mot}),
solving for ${\bd R}^{\prime\prime}$, yields

\beqa
{\bd R}^{\prime\prime} & = & \frac{c^2}{6} (3\xi_1 - \xi_0 )R\sigma_-
- \frac{4\pi\kappa}{3} (1+\frac{3\alpha}{c^2}) \rho_0
{\bd R}^{-(2+\frac{3\alpha}{c^2})}
~~~.
\nonumber \\
\label{rpp}
\eeqa
We will also need the relation

\beqa
\frac{dlnV_-}{dt} & = & \frac{1}{V_-}\frac{dV_-}{dt}
~=~ \frac{1}{{\bd R}_-^3} \frac{d}{dt} {\bd R}_-^3
~=~ \frac{3{\bd R}_-^\prime}{{\bd R}_-}     \nonumber \\
& = & \frac{d{\rm ln}{\bd R}_-^3}{dt}
~~~.
\label{dlnv}
\eeqa

Now we remember our former result that $\xi_1$ has to be {\it approximately}
equal to $\xi_0$ (Eq. (\ref{bedi})).
Due to this we can assume that the following relation also holds
approximately:

\beqa
\xi_1 & = & \beta\xi_0
~~~,
\label{60}
\eeqa
where $\beta$ is an additional parameter of the theory, describing the
deviation from a constant Hubble parameter $H$. In principle,
one can also use a power expansion of $\xi_1$ in terms of $\xi_0$, which
would only introduce more parameters. 
The $\beta$ will later be related to
observable quantities, like the Hubble constant and the deceleration
parameter.
Eq. ({\ref{60}) gives us the missing condition,
with the prize of having to introduce an additional parameter. Another
possibility is to use the approximate expression of $\xi_0$ in terms
of the ratio $({\bd R}_r^\prime /{\bd R}_r)$, which gives
$\xi_0 = (3/c^2)H$, and use experimental observations for $H$.

Using (\ref{eq17}), we obtain for the differential equation for
$\xi_0$

\beqa
\frac{d\xi_0}{dt} & = & (\beta -1)\frac{d({\rm ln}{\bd R}_-^3)}{dt}\xi_0
\nonumber \\
& = & \frac{d({\rm ln}{\bd R}_-^{3(\beta -1)})}{dt}\xi_0
~~~,
\eeqa
with the solution

\beqa
\xi_0 & = & \Lambda {\bd R}_-^{3(\beta - 1)}
~~~.
\label{eq67}
\eeqa
This leaves us with the two, yet undetermined, parameters
$\Lambda$ and $\beta$. There are many different scenarios:\\
i) $\beta = 1$: Then $\xi_1 = \xi_0= \Lambda$ is constant.
\\
ii) $\beta\ne 0$: This will lead (see further below) to de-accelerated and
accelerated systems, depending on the value of $\beta$. Also the
acceleration as a function of the radius of the universe
(which can be correlated to time of evolution) depends on $\beta$
and $\Lambda$.

The real part of (\ref{rpp}) is obtained by $R_r^{\prime\prime}$
= $\frac{1}{2}\left( R_+^{\prime\prime} + R_-^{\prime\prime} \right)$.
Because the minimal length scale $l$ is extremely small, we can assume that 
${\bd R}_+\approx{\bd R}_-\approx{\bd R}_r$. The real part of
(\ref{rpp}) is obtained by summing the $\sigma_+$ and $\sigma_-$ components
and dividing the result by 2. 
We will also assume that $\alpha_+\approx \alpha_- = \alpha$, which is
reasonable because the $\alpha$ relates the pressure and the density, which
are both pseudo-real.
Using also (\ref{60}) and (\ref{eq67}) 
gives the final form of the equation of motion for the radius
of the universe

\beqa
{\bd R}_r^{\prime\prime} & = & \frac{c^2}{12} (3\beta - 1)\Lambda
R_r^{3(\beta - 1)+1} \nonumber \\
&& - \frac{4\pi\kappa}{3} (1+\frac{3\alpha}{c^2}) \rho_{0}
{\bd R}_r^{-3(1+\frac{\alpha}{c^2})+1}
~~~.
\label{rpp-2}
\eeqa
We have assumed that the density is real.
The first term comes from the $\xi$-functions.

\section{Consequences}

In this section we shall discuss the consequences of the important result (\ref{rpp-2}).

When $\beta = 1$
(cosmological constant), the sign of the first term in
(\ref{rpp-2}) is positive and contributes to the acceleration of
the universe. The acceleration increases with the radius of the universe.
For a general $\beta$,
the acceleration is positive, as long as $\beta > \frac{1}{3}$, it is
negative (deceleration) for $\beta < \frac{1}{3}$. For
$\beta = \frac{1}{3}$ no additional
acceleration nor deceleration takes place.
The last term in (\ref{rpp-2}) is
always negative, i.e., it represents a contribution which contributes to
the deceleration of the universe. The behavior of how the accelerating term
behaves as a function in ${\bd R}_r$ is also determined by $\beta$. If the
exponent of ${\bd R}_r$ is positive, the acceleration increases with
${\bd R}_r$, if $\beta > \frac{2}{3}$, while it decreases with
${\bd R}_r$ for $\beta < \frac{2}{3}$.

The solution (\ref{rpp-2}) leaves space for a number of different possible
scenarios.
{\it In order to proceed, we will make the following assumption}:
For simplicity, we assume as before that
the parameter $\alpha$ and the density $\rho_0$ are pseudo-real, i.e.,
$\alpha_+ = \alpha_- = \alpha$ and $\rho_{0+}=\rho_{0-}=\rho$.
In this case the solution simplifies to

\beqa
{\bd R}_r^{\prime\prime} & = & \frac{c^2}{12} (3\beta - 1)\Lambda
R_r^{3(\beta - 1)+1} \nonumber \\
&& - \frac{4\pi\kappa}{3} (1+\frac{3\alpha}{c^2}) \rho_{0}
{\bd R}_r^{-3(1+\frac{\alpha}{c^2})+1}
~~~.
\label{rpp-3}
\eeqa
With no dark energy, $\Lambda=0$, using (\ref{eq67}), we have
$\xi_1 = \xi_0 = 0$ and this equation reduces to the one in
\cite{adler}.

Let us now discuss several particular values of $\beta$. For that purpose
we define

\beqa
{\widetilde \Lambda} & = & \frac{c^2}{16\pi \kappa }
\frac{\Lambda}{\rho_0}
~~~.
\label{lam-til}
\eeqa
Note that according to (\ref{rho-lambda}) the $\Lambda$ is proportional to
$\rho_\Lambda$, the density of the dark energy, with the same
proportionality factor. Here $\rho_0$ is the mass
density of the universe. Both densities are of the same order, implying that
${\widetilde \Lambda}$ is of the order of 1.

We can now rewrite (\ref{rpp-3}) into

\beqa
{\widetilde {\bd R}}_r^{\prime\prime} & = &
\frac{{\bd R}_r^{\prime\prime}}{\left(\frac{4\pi\kappa}{3}\right)\rho_0}
\nonumber \\
& = & {\widetilde \Lambda}\left(3\beta -1\right) {\bd R}_r^{3(\beta - 1)+1}
- \left(1+\frac{3\alpha}{c^2}\right){\bd R}_r^{-3(1+\frac{\alpha}{c^2})+1}
\label{eq71}
~~~.
\eeqa

In what follows, we discuss the case of dust dominated universe,
i.e., $\alpha = 0$. For the
case of a relativistic ideal gas $\alpha = \frac{1}{3}$, the
results show the same characteristics.
We will take arbitrarily different values of $\beta$, which are chosen such
that we will have the case of the cosmological {\it constant} $\Lambda$,
a case which will represent the solution of the big-rip-off and two new
solutions. These solutions are not necessarily represented in nature, i.e., $\beta$ might have a different intermediate value as those in the examples.
With this we get for

\noindent
{\bf a) $\beta = 1$}: ($\xi_1=\xi_0=\Lambda$) \\

\beqa
{\widetilde {\bd R}}_r^{\prime\prime} & = & 2{\widetilde \Lambda}
{\bd R}_r - {\bd R}_r^{-2}
~~~.
\eeqa
The universe is accelerated by the first contribution and decelerated by
the second one. For small ${\bd R}_r$ the universe is decelerated.
For large ${\bd R}_r$ the first term starts to dominate and the universe
is from then on accelerated. The turning point is at

\beqa
{\bd R}_r & \approx & 0.79/{\widetilde \Lambda}^{\frac{1}{3}} ~~~.
\eeqa
If we set the radius of today at ${\bd R}_0 = 1$, a common definition
of scale for the present epoch, the result implies that for ${\widetilde \Lambda}=1$
acceleration did set in after the
universe passed 80 percent of its radius. This case corresponds to a constant
cosmological function $\Lambda$.

\noindent
{\bf b) $\beta = \frac{4}{3}$}: \\

\beqa
{\widetilde {\bd R}}_r^{\prime\prime} & = &
3{\widetilde \Lambda}{\bd R}_r^{2} - {\bd R}_r^{-2}
~~~.
\eeqa
In this case, the acceleration increases with the second power in
${\bd R}_r$, stronger than only with a cosmological constant.
The $\xi_0$ function (\ref{eq67}) is then given by $\Lambda{\bd R}_-$, i.e., the
dark energy density, represented by $\xi_0$, increases with the radius of the
universe..
This is like
the big rip-off, which is discussed in the literature.
The break-even point, i.e. when acceleration is equal to deceleration,
is reached for

\beqa
{\bd R}_r & = & 1/(3{\widetilde \Lambda})^{\frac{1}{4}} ~~~.
\eeqa
For ${\widetilde \Lambda}=1$ the break-even point is reached when the universe is
$1/3$ of its present radius, thus, earlier than in case a).

\noindent
{\bf c) $\beta = \frac{1}{2}$}:
Remember that $\alpha=0$ (dust dominated
universe)! Then, from (\ref{eq71}) we get
\\

\beqa
{\widetilde {\bd R}}_r^{\prime\prime} & = &
\frac{1}{2}{\widetilde \Lambda}{\bd R}_r^{-\frac{1}{2}} - {\bd R}_r^{-2}
~~~.
\label{max}
\eeqa
In this situation, the dark energy behaves as (use Eq. (\ref{eq67}))
$\rho_0=\frac{\Lambda}{{\bd R}_r^{3/2}}$, i.e., the density of the dark
energy decreases with the radius (time) of the universe.

{\it This is really a new solution}!
The accelerating and the decelerating parts are decreasing with the size
of the universe, {\it but at a different rate}. For small ${\bf R}_r$,
the second term dominates and the universe is decelerated, while for
sufficient large ${\bd R}_r$ the first, accelerating, term dominates
and the universe is accelerated! The break-even point is at

\beqa
{\bd R}_r & \approx & 2^{\frac{2}{3}}/{\widetilde \Lambda}^{\frac{2}{3}} ~~~,
\eeqa
i.e., for ${\widetilde \Lambda}=1$
the universe at this break-even point
will be at about $2^{\frac{2}{3}}\approx 1.59$ times of its present radius.
Of course, this can be changed using different values of
${\widetilde \Lambda}$. For ${\widetilde \Lambda}=3$ the break-even point is at
76 percent of the radius of the universe.
This case is plotted in figure 1. The universe starts to be accelerated
after having reached ${\bd R}_r=0.76$ (units in ${\bd R}_{r0}$).
After that the acceleration increases.
However, having reached the radius ${\bd R}_r \approx 1.9$,
i.e., nearly twice the actual radius of the universe, it
reaches a maximum and after that the acceleration is {\it decreasing},
reaching asymptotically zero. This universe will never collapse but
reach an asymptotically non-accelerating state

\begin{figure}[ph]
\centerline{\epsfxsize=10cm\epsffile{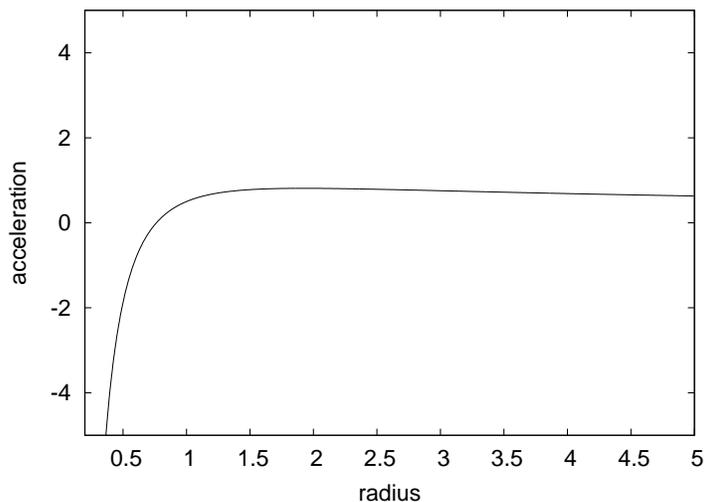}}
\caption{
Dependence of the scaled acceleration as a function of ${\bd R}_r$,
the radius of the universe, for $\beta = \frac{1}{2}$.
In this figure ${\widetilde \Lambda} = 3$.
The maximum of this function can be deduced from (\ref{max}),
giving
${\bd R}_{{\rm max}}= \left( 8/{\widetilde \Lambda} \right)^{\frac{2}{3}}$
= 1.923. The maximum can be barely seen in the figure due to the
extreme slow decrease of the function.
The question is also: Where are we now?
Before or after the maximum?
}
\label{fig1}
\end{figure}

\noindent
{\bf d) $\beta = \frac{2}{3}$}:
Remember that $\alpha=0$ (dust dominated
universe)! Then, from (\ref{eq71}) we get
\\

\beqa
{\widetilde {\bd R}}_r^{\prime\prime} & = &
{\widetilde \Lambda} - {\bd R}_r^{-2}
~~~.
\eeqa
{\it This is also a new solution}. The break-even point is now at

\beqa
{\bd R}_r & \approx & 1/\sqrt{{\widetilde \Lambda}} ~~~.
\eeqa
This solution is also special in the sense
that the asymptotic acceleration of the universe {\it is constant}
(${\bd R}_r^{\prime\prime} = {\widetilde \Lambda}$). Using (\ref{eq67})
leads to the dependence
$\rho_0=\Lambda / {\bd R}_r$
of the dark energy on the radius of the universe, i.e., it also
decreases with the radius (time) of the universe.

\begin{figure}[ph]
\centerline{\epsfxsize=10cm\epsffile{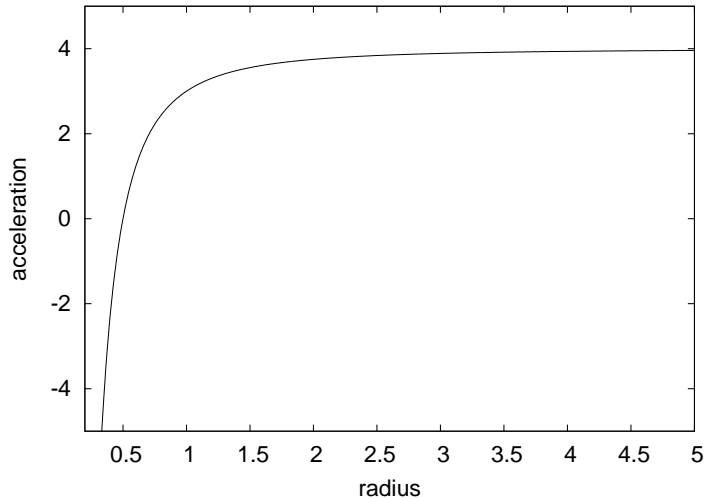}}
\caption{
Dependence of the scaled acceleration as a function in ${\bd R}_r$,
the radius of the universe, for $\beta = \frac{2}{3}$. In this figure,
the ${\widetilde \Lambda}=4$.
}
\label{fig2}
\end{figure}

In all cases ${\widetilde \Lambda}$ can be fitted to the observation
at which stage
the net acceleration did set in, overcoming the deceleration term
in (\ref{rpp-3}). {\it The new part here} is that
other solutions exist than the standard ones: \\
i) There is the possibility of a constant asymptotic acceleration. \\
ii) In another solution, the expansion of the universe, after its decelerating period,
gets accelerated. The accelerations reach a maximum and vanish
asymptotically. In this case the universe approaches, for large times,
an ever expanding, non-accelerating phase. \\
iii) In all cases, the universe is first decelerated and after a so-called
{\it break-even point} it starts to accelerate. \\
iv) Of course, all standard solutions are obtained (cosmological
constant and rip-off).
\\
In order to calculate numerically observable consequences, we have to
know the exact form of the $\xi_k$ functions, which we were unable to
deduce from first principles.
One possibility is to use the calculated distribution of dark
energy, as for example done in \cite{mich1}.
An alternative is to use the parametrization given in (\ref{60}).
This implies the use of an additional parameter ($\beta$) and is equivalent
to known considerations in the literature \cite{peebles}.
The acceleration in each solution is a consequence of the $\xi_k$ functions.
As discussed above, {\bf they represent contributions to the energy-momentum tensor,
equivalent to the dark energy. In the model considered, this dark energy
is in general not a constant but may vary in time, i.e., with the radius of
the universe}.

\begin{center}
{\bf Extraction of $\beta$:}
\end{center}
We can try to connect the value of $\beta$ to observable quantities. For that,
we start from Eq. ({\ref{estimation}), without the approximation in the
last line. We get

\beqa
\frac{{\bd R}_r^{\prime\prime}}{{\bd R}_r} & = & \left[ \frac{H^\prime}{H}
+ \frac{{\bd R}_r^\prime}{{\bd R}_r} \right]
\frac{{\bd R}_r^\prime}{{\bd R}_r}
\nonumber \\
& = & \left[ \frac{H^\prime}{H} + H \right] H ~=~ H^\prime + H^2
~~~.
\eeqa
Substituting this into the expression for $\xi_1$ (Eq. (\ref{x0-x1})),
setting $k=0$, we obtain

\beqa
\xi_1 & = & \frac{1}{c^2} H^2 + \frac{2}{c^2}\left( H^\prime + H^2 \right)
\nonumber \\
& = & \frac{3}{c^2}H^2 + \frac{2}{c^2}H^\prime
\nonumber \\
& = & \xi_0 + \frac{2}{c^2}H^\prime
\nonumber \\
& = & \beta \xi_0
~~~,
\eeqa
where we have used that $\xi_1 = \beta\xi_0$.
Using $\xi_0 = \frac{3}{c^2}H^2$ (see (\ref{bedi}))
and solving for $\beta$, we obtain
the final result

\beqa
\beta & = & 1+ \frac{2}{3}\frac{H^\prime}{H^2}
~~~.
\label{eq87}
\eeqa

We obtained further above that for $\beta > \frac{1}{3}$ the universe
is accelerated after a given radius. This corresponds to $\frac{H^\prime}{H^2}>-1$.
In order to get deeper insight, we use the deceleration parameter,
which is a measure whether the universe is accelerated or decelerated, depending
on the sign of this parameter.
The deceleration parameter is defined as \cite{adler}
$q=-\frac{{\bd R}_r^{\prime\prime}{\bd R}_r}{{\bd R}_r^{\prime 2}}$,
which gives 
$q=-\left[1+\frac{H^\prime}{H^2}\right]$.
The universe is accelerated when
$\beta < \frac{1}{3}$, or $\frac{H^\prime}{H^2}<-1$, or $q>0$.
The acceleration increases with ${\bd R}_r$ when $\beta > \frac{2}{3}$,
or $\frac{H^\prime}{H^2}>-\frac{1}{2}$, or $q<-\frac{1}{2}$.

In conclusion, a measurement of the change of the Hubble constant with time
will lead to a determination of the parameter $\beta$ as a function
of time. Though, the last
considerations clarify the role of $\beta$, we are suffering still by
the problem that we have to know the solution of
$H=\frac{{\bd R}_r^\prime}{{\bd R}_r}$.
This can be done up to now only through the experimental measurement of the
Hubble parameter $H$.

\begin{center}
{\bf A model including dust and radiation, k=0:}
\end{center}

Up to now, we did only consider one density component (dust or radiation)
and the pseudo-complex contribution, given by the $\xi_k$ functions.
Realistic models involve both components, as can be seen in \cite{peebles}.
Expressing the ratio of the radii ${\bd R}_{r0}$ and ${\bd R}_r$, the present radius
of the universe and the one at a redshift $z$, respectively, in terms
of the redshift, we obtain \cite{adler,peebles}

\beqa
\frac{{\bd R}_{r0}}{{\bd R}_r} & = & (1+z)
~~~,
\eeqa
where the index $r$ refers to the real value of the radius. We obtain for
the ratio of the velocity and the the radius of the universe \cite{peebles}

\beqa
\frac{{\bd R}^\prime_{r}}{{\bd R}_r} & = & H ~=~
H_0^2 \left\{ \Omega_d (1+z)^{3} + \Omega_r (1+z)^{4} + \Omega_{\Lambda}
f(z) \right\}
\nonumber \\
~~~
\eeqa
where the index $d$ refers to the dust part and the index $r$ to the
radiation part. We do not include the contribution due to $k\neq 0$, because
we consider a flat universe.
The factor $H_0^2$ is the square
of the present Hubble constant.  Using our previous result,
the function $f(z)$ is given by

\beqa
f(z) & = & (1+z)^{3(\beta - 1)} ~=~ (1+z)^{3(1+w)}
~~~,
\eeqa
($\beta = 2 + w$)
where we made a connection to the notation used in \cite{peebles}. Our
result states that when the Hubble constant changes in time, there
{\it must} be a deviation from $\beta=1$, which corresponds to
$w=-1$, the case of a cosmological constant, constant in time.
The deviation from $\beta = 1$ cannot be large when $H^\prime$,
the time derivative of the Hubble constant, is small.
Up to now, we only find a parametrization of the $\xi_k$ functions
in terms of the Hubble parameter or the parameter $\beta$. The deeper
origin of the value of $\xi_k$ can probably explained by fundamental
theories like string theories.

Note, that the relation $\beta = 2+ w$ with (\ref{eq87}) gives a relation
of $w$ to the Hubble parameter and its derivative in time, which makes
definite predictions on $w$, once $H$ and $H^\prime$ are known. To our
knowledge, this is not presented elsewhere.

\section{Conclusions}

We have applied the pseudo-complex formalism to extend the Robertson-Walker
model of the universe to the {\it pc}-RW model. The main results are that \\
1) The model introduces automatically a contribution which
is equal to the cosmological constant
or dark energy which may depend on the radius of the universe. \\
2) The cosmological "constant" is a constant
when the Hubble constant is constant too. When the Hubble
constant changes slightly with time, our model predicts deviations from
the cosmological constant, depending on the redshift (time of expansion).
The amount of deviations depends on the exact form of
the $\xi_k$ functions.
\\
3) The deviation can be obtained, once the radius of the universe,
as a function of time, is known. Within our theory, we obtain several
possible dependencies of the dark energy as a function in the radius of the
universe, depending on the parameter $\beta$.
\\
4) We also obtained several possible evolutions of the universe. Besides
the solution of a constant dark energy density and the rip-off scenario, we
also obtained solutions where the acceleration tends for infinite time
towards zero or a constant value (see Figs. \ref{fig1} and \ref{fig2}). \\
5) We obtained a relation between $w=\beta-2$ and $H$ and $H^\prime$.
Once $H$ and $H^\prime$ are known, the $w$ value can be deduced.

The origin of the $\xi_k$ functions might have a deeper microscopic origin,
which we do not explore here. Probably, only such a deeper microscopic
understanding will fix the dependence of $\xi_k$ on the radius of the
universe (see for example \cite{mich1}).
Nevertheless, the classical picture presented here enlightens
and simplifies the description of different possible
evolution scenarios of the universe.

We have not yet investigated the role of the minimal length parameter $l$,
which also appears in the pseudo-complex formulation. In field theory
its function is to render the theory regularized \cite{hess2}. We suspect
that this also happens in the pseudo-complex formulation of General
Relativity and might give a hint on how to quantize this theory.
In a future publication we intend to investigate the role of the minimal
length scale $l$.

We saw that the modified variational principle
$\delta S~ \epsilon~ {\bf P}^0$ has important consequences as the appearance
of dark energy. It also provides a simpler description of effects of
the dark energy, obtained via quite involved numerical calculations, as
for example in \cite{mich1}.
These features are a hint that the variational
principle has to be probably modified as proposed.

\section*{Acknowledgments}
P.O.H. wants to express sincere gratitude for the possibility to work at
the {\it Frankfurt Institute of Advanced Studies} and of the excellent
working atmosphere encountered there. He also acknowledges financial
support from DGAPA and CONACyT.

\end{document}